\documentclass[%
 aip,
 amsmath,amssymb,
 reprint,%
]{revtex4-1}

\usepackage{graphicx}% Include figure files
\usepackage{dcolumn}% Align table columns on decimal point
\usepackage{bm}% bold math
\usepackage[utf8]{inputenc}
\usepackage[T1]{fontenc}
\usepackage{mathptmx}
\usepackage{etoolbox}

\makeatletter
\def\@email#1#2{%
 \endgroup
 \patchcmd{\titleblock@produce}
  {\frontmatter@RRAPformat}
  {\frontmatter@RRAPformat{\produce@RRAP{*#1\href{mailto:#2}{#2}}}\frontmatter@RRAPformat}
  {}{}
}%
\makeatother
\begin{document}

\preprint{arXiv}

\title{A simple method for programming and analyzing multilevel crystallization states in phase-change materials thin films}
%\title{Programming and analyzing multilevels of partial crystallization in phase-change materials thin films}
%Alternative titles:
% A simple method for programming and analyzing partial crystallization in phase-change materials thin films
%Real-time monitoring of multilevel crystallization
% Force line breaks with \\

\author{Arnaud Taute}%
\affiliation{Univ Lyon, CNRS, ECL, INSA Lyon, UCBL, CPE, INL UMR5270, 69134 Ecully, France}%
\affiliation{STMicroelectronics, 850 Rue Jean Monnet, Crolles, 38920, France}

\author{Sadek Al-Jibouri}%
\affiliation{Univ Lyon, CNRS, ECL, INSA Lyon, UCBL, CPE, INL UMR5270, 69134 Ecully, France}%

\author{Capucine Laprais}%
\affiliation{Univ Lyon, CNRS, ECL, INSA Lyon, UCBL, CPE, INL UMR5270, 69134 Ecully, France}%

\author{Stéphane Monfray}%
\affiliation{STMicroelectronics, 850 Rue Jean Monnet, Crolles, 38920, France}

\author{Julien Lumeau}%
\affiliation{Aix Marseille Univ, CNRS, Centrale Marseille, Institut Fresnel, F-13013 Marseille, France}

\author{Antonin Moreau}%
\affiliation{Aix Marseille Univ, CNRS, Centrale Marseille, Institut Fresnel, F-13013 Marseille, France}

\author{Xavier Letartre}%

\author{Nicolas Baboux}%

\author{Guillaume Saint-Girons}%

\author{Lotfi Berguiga}%

\author{Sébastien Cueff}%
\email{sebastien.cueff@cnrs.fr}

\affiliation{Univ Lyon, CNRS, ECL, INSA Lyon, UCBL, CPE, INL UMR5270, 69134 Ecully, France}%

\date{\today}% It is always \today, today,
             %  but any date may be explicitly specified

\begin{abstract}
We propose and demonstrate a simple method to accurately monitor and program arbitrary states of partial crystallization in phase-change materials (PCMs). The method relies both on the optical absorption in PCMs as well as on the physics of crystallization kinetics. Instead of raising temperature incrementally to increase the fraction of crystallized material, we leverage the time evolution of crystallization at constant temperatures and couple this to a real-time optical monitoring to precisely control the change of phase. We experimentally demonstrate this scheme by encoding a dozen of distinct states of crystallization in two different PCMs: GST and Sb$_2$S$_3$. We further exploit this 'time-crystallization' for the in-situ analysis of phase change mechanisms and demonstrate that the physics of crystallization in Sb$_2$S$_3$ is fully described by the so-called Johnson-Mehl-Avrami-Kolmogorov formalism. The presented method not only paves the way towards real-time and model-free programming of non-volatile reconfigurable photonic integrated devices, but also provides crucial insights into the physics of crystallization in PCMs.

\end{abstract}

\maketitle

Phase change materials (PCMs) are currently revolutionizing nanophotonics by providing ways to tune and reconfigure optical functionalities without any moving parts. The rapid rearrangement of atoms at the nanoscale translates into very large modifications of optical properties. Building on this phenomenon, the last decade has witnessed many exciting reports of novel devices exploiting PCMs such as for example beam-steering, tunable light emission, reflection and absorption, programmable metasurfaces and reconfigurable neural networks \cite{wuttig2017phase,abdollahramezani2020tunable,cueff2020vo2,de2018nonvolatile,cueff2015dynamic,feldmann2019all,feldmann2021parallel, tripathi2021tunable, fang2022ultra, zhang2021electrically, moitra2022programmable}. A large majority of the first studies were using PCMs as simple binary on-off switches, in which the ON state is the amorphous phase and the OFF state the fully crystalline phase. However, PCMs present another degree of freedom for tunability: the possibility to encode multilevel non-volatile states via partial crystallization. Recently, several works exploited this potential of PCMs for devices in waveguides, thin-films or metasurfaces \cite{rios2015integrated,cueff_reconfigurable_2021,mikheeva2022space, li2019fast, farmakidis2022electronically}. 
Even though it may appear straightforward to prepare a thin-film to a desired level of partial crystallization – after all, one should just bring it to the correct temperature for a given duration – several issues make it a serious challenge to face. The main reason lies in the abruptness of the crystallization process, which may occur within nanoseconds after overcoming the temperature threshold for crystallization. Moreover, monitoring the progressive crystallization of a thin-film usually requires either sequential processes with destructive analysis (micrograph analysis of alloys or transmission electron microscopy) or spectral measurements whose acquisition time and analysis are typically longer than the timescale of crystallization (X-ray diffraction, Raman or spectroscopic ellipsometry). Neither of these techniques are compatible with a real-time analysis of the crystalline fraction of a thin-film. In addition, most of the PCMs undergo a mechanical contraction (reduction of thickness) upon crystallization. This contraction could be as high as 20\% and further complicates the analysis of crystallization fraction, given thickness and complex refractive index are correlated parameters when measured optically.
However, both fundamental studies and future applications call for precise and rapid methods to analyze and program the crystalline fraction in PCMs thin-films.

Here, we propose a reliable, yet very simple method to program PCMs thin-films at a desired arbitrary crystalline fraction by following its state in real-time. To do so, we exploit both the optical absorption and the time-dependence of crystallization at temperatures near the threshold of crystallization. Additionally, we show that this method provides a way for the in-situ analysis of the crystallization mechanisms at play in PCM thin-films.

\begin{figure*}[htp!]
\centering\includegraphics[width=1\linewidth]{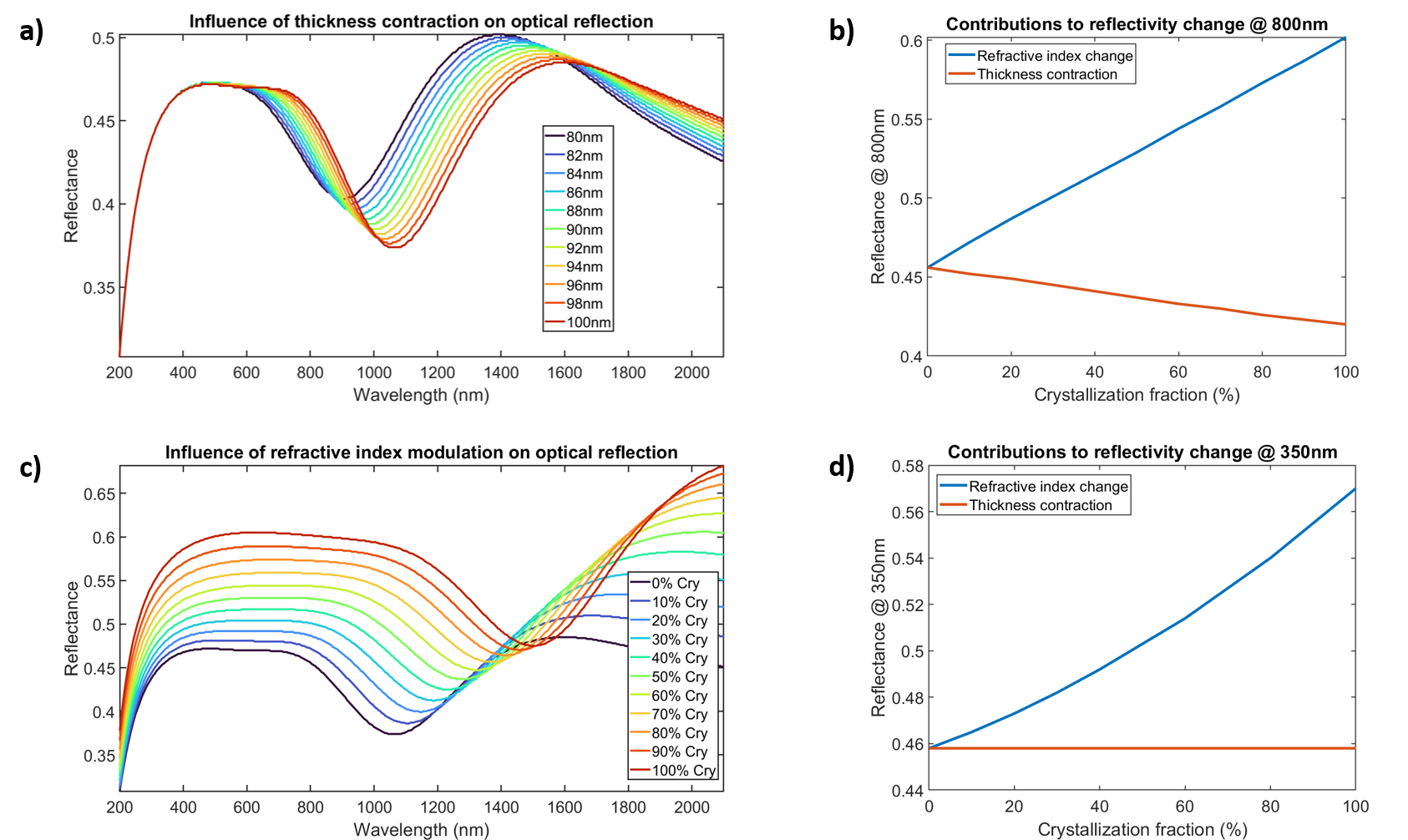}
\caption{Theoretical reflectivity of a 100-nm-thick GST film on silicon substrate at normal incidence a) Influence of a thickness contraction of 20\% on the optical reflection and c) Influence of a refractive index change corresponding to the crystallization of the GST thin film. b) and d) Respective contributions of thickness contraction and refractive index modulation to reflectivity change at wavelengths of b) 800nm and d) 350 nm}
\label{fig1}
\end{figure*}

A standard method of analyzing the optical properties of PCMs in their different phases is to use spectroscopic ellipsometry coupled with a heat cell. Each state corresponding to a partial crystallization of the PCM is fitted to an oscillator model and the associated dispersion is then extracted (see e.g. \cite{cueff_reconfigurable_2021,gutierrez2022characterizing}). However, the analysis is far from being straightforward as the PCMs usually go through a contraction upon their change of phase. It follows that two correlated parameters are simultaneously varying during this transition: the complex refractive index and the thickness.
To illustrate how this double variation can seriously affect the analysis, we consider the normal incidence reflection of a 100-nm-thick Ge$_2$Sb$_2$Te$_5$ (GST) layer on a silicon substrate. We model the optical dispersion of the amorphous and crystalline phase using a Tauc-Lorentz model optimized from previous measurements \cite{cueff_reconfigurable_2021} and the intermediate phases are modelled using a Bruggeman effective medium approximation mixing amorphous and crystalline phases in varying proportions (see Supplemental Material for more information). In Figure \ref{fig1} a) and b), we show the separate contributions of thickness reduction and refractive index modulation -- both due to a progressive crystallization -- on the overall measured reflection. From the results displayed, it is clear that the two parameters affect differently the reflection and set challenges for the analysis, given thickness and refractive index are correlated parameters for optical measurements.

In most cases, this correlation can be lifted by adding another independent measurement to untangle the two contributions. For example, X-ray reflectivity can be used to precisely measure the thickness of the amorphous and crystalline phases and the measured values of thicknesses can then be fixed and no longer used as a free fit parameter in the ellipsometric models.
Unfortunately, when one needs to follow crystallization in situ and in real-time this method is not suitable.

However, a close inspection to the reflection variations in Figure \ref{fig1} reveals a spectral region, below 500 nm, at which the reflection is insensitive to thickness modifications. The physical reason of this phenomenon is due to the optical absorption of GST: this spectral region falls above the bandgap of GST, which results in total absorption of light before it reaches the substrate. On the contrary, at the same wavelengths the reflection is strongly affected by refractive index variations. We therefore have a spectral region in which we can easily untangle the two correlated parameters and measure the sole contribution of refractive index modulations.

Following the crystallization in real time could thus be done by isolating conditions of optical measurements that are only sensitive to refractive index change and not to thickness modifications. However, crystallizing PCMs requires heating the materials above their crystallization temperatures. When dealing with temperature variations, there is another important factor to take into account: each materials present their own specific thermo-optic coefficients, resulting in non-straightforward modulations of optical properties as a function of temperature in multi-materials stacks.
To circumvent these two additional sources of modulation (contraction and thermo-optical effects), we apply two simple rules: i) we measure the variation of optical properties at a fixed wavelength where the absorption in the PCM thin film is high enough to ensure the measured variations only depend on the refractive index of the PCM layer. ii) Instead of incrementally raising the temperature to progressively crystallize the PCM, we set the temperature to a fixed value close to the onset of crystallization and let the PCM crystallize as a function of time. Indeed, crystallization is based on the diffusion and re-arrangement of atoms through a temperature-activated process that typically follows an Arrhenius law. The diffusion length of atoms both depends on the diffusion coefficient and time: $l=\sqrt{Dt}$. Hence, progressive crystallization of materials can be produced at constant temperature. By following these guidelines, we rule out all parasitic contributions to optical modulations and ensure that the measured optical variations solely come from refractive index variation caused by the progressive crystallization of the PCM thin film.

\begin{figure*}[!ht] 
\centering
   \includegraphics[width=1\linewidth]{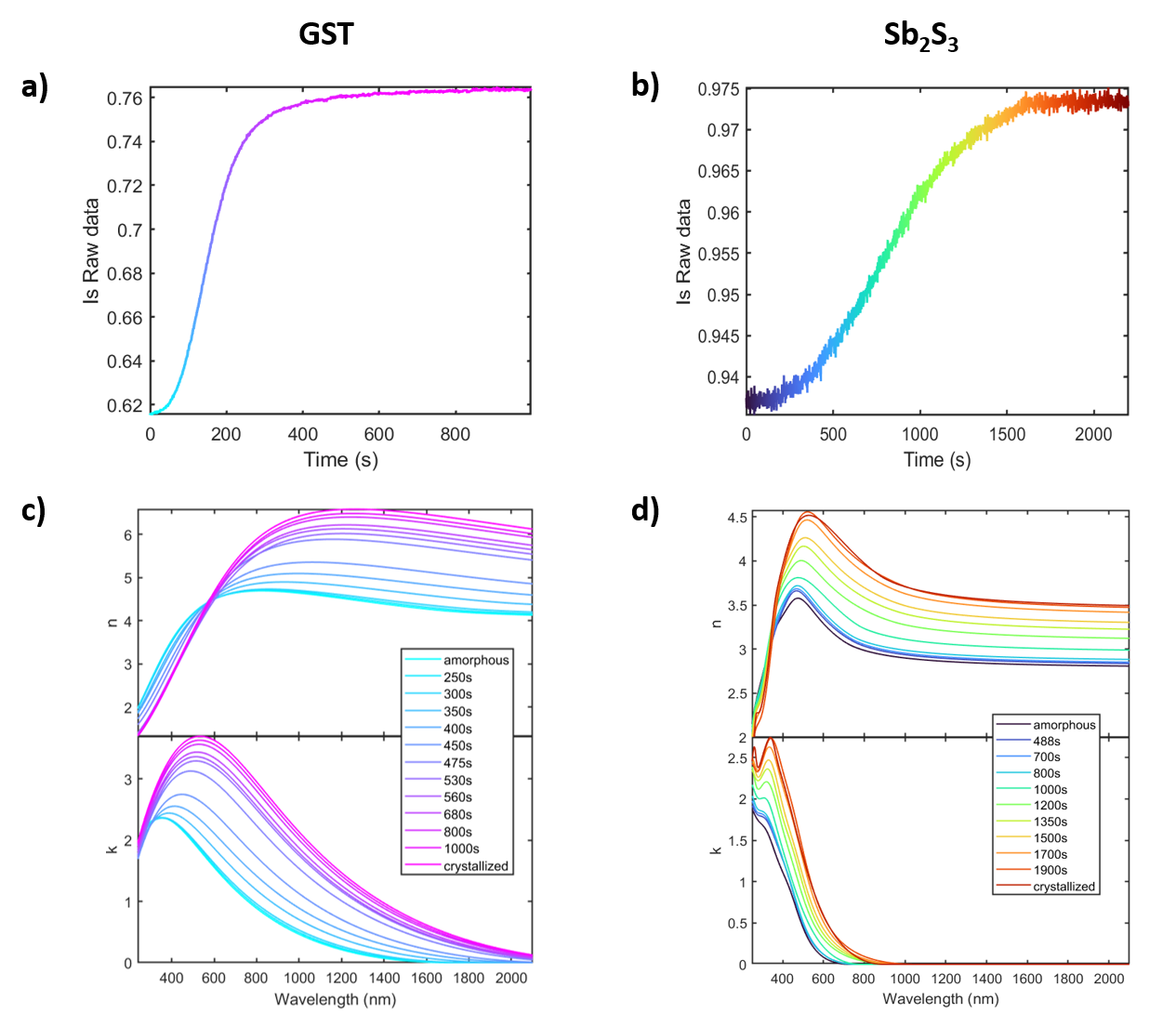}
  \caption{Variation of the Is parameter as measured by ellipsometry versus time for a) GST and b) Sb$_2$S$_3$. For each PCM, the heating was stopped at different times and ellipsometry spectra were recorded at room temperature, resulting in the programming of multiple states of partial crystallization, as shown in the extracted dispersions for c) GST and d) Sb$_2$S$_3$.}
 \label{Figure2}
\end{figure*}

We experimentally demonstrate this principle with two different PCMs: a 47-nm-thick GST layer deposited on silicon and capped with a 5-nm-thick SiO$_2$ layer, and an 180-nm-thick Sb$_2$S$_3$ layer on silicon capped with 55nm of SiO$_2$. For the optical monitoring, we use an ellipsometer at a fixed incidence angle of 70° and measure the reflection at a single wavelength as a function of time for different temperatures, using a Linkam heat-cell with controlled temperature. The wavelengths and temperatures were set at 500 nm and 145°C for the GST sample and 435 nm and 260°C for the Sb$_2$S$_3$ sample. These temperatures were chosen such that the modifications of the measured parameters are slow enough to be easily monitored in time. Before reaching these values, the temperature was progressively raised, to ensure the thermal equilibrium is reached on the sample as soon as the crystallization temperature is set. As displayed in figure \ref{Figure2} a) and b), we can see that the measured raw $Is$ parameter (which is a combination of the $\Delta$ and $\Psi$ ellipsometric angles -- see supplemental material for more details) follows an exponential trend as a function of time and tends to an asymptotic value for both layers.

As we have ruled out any other contributions to the modulation of the optical properties, the measured variations solely correspond to refractive index change of the PCMs through its progressive crystallization. To demonstrate this, we stop the crystallization by simply removing the sample from the heat cell at various times, measure the ellipsometry spectra of the sample at room temperature and fit the measurements to a Tauc-Lorentz oscillator model (more details in the supplement). As displayed in fig. \ref{Figure2} c) and d), we see that the progressive multilevel crystallization is confirmed and that we were able to program the states of both PCM films to more than ten different intermediate values of refractive index between the amorphous and crystalline ones. We also directly confirm here that all intermediate partially crystallized states of Sb$_2$S$_3$ retain the ultralow optical absorption in the near-infrared region \cite{dong2019wide,delaney2020new,fang2021non,gutierrez2022interlaboratory}

This is the first major outcome of this report: by carefully selecting conditions for progressive crystallization, associated with real-time monitoring, we provide a simple and robust method to program a PCM thin-film in an arbitrary state of partial crystallization. This scheme only requires a light source with appropriate wavelength, a detector and a proper initial calibration, but neither require complicated optical setups with spectrometers and filters nor a model and associated fits for the analysis.

In addition to providing a simple means to program the crystallization fraction in the PCM layers, this method also enables more fundamental insights into the crystallization mechanisms of PCMs.
As described in the pioneering works of Johnson, Mehl, Avrami and Kolmogorov \cite{johnson1939rf,avrami1939kinetics,avrami1940kinetics,avrami1941granulation,kolmogorov1937static}, change of phase proceeds via an active growth of transformation initiated around a nuclei. The volume of transformed material follows the so-called Johnson-Mehl-Avrami-Kolmogorov (JMAK) equation:

\begin{equation}
Y(t)=1-e^{(-q.t)^m}   
\end{equation}

\begin{figure*}[!ht] 
\centering
   \includegraphics[width=1\linewidth]{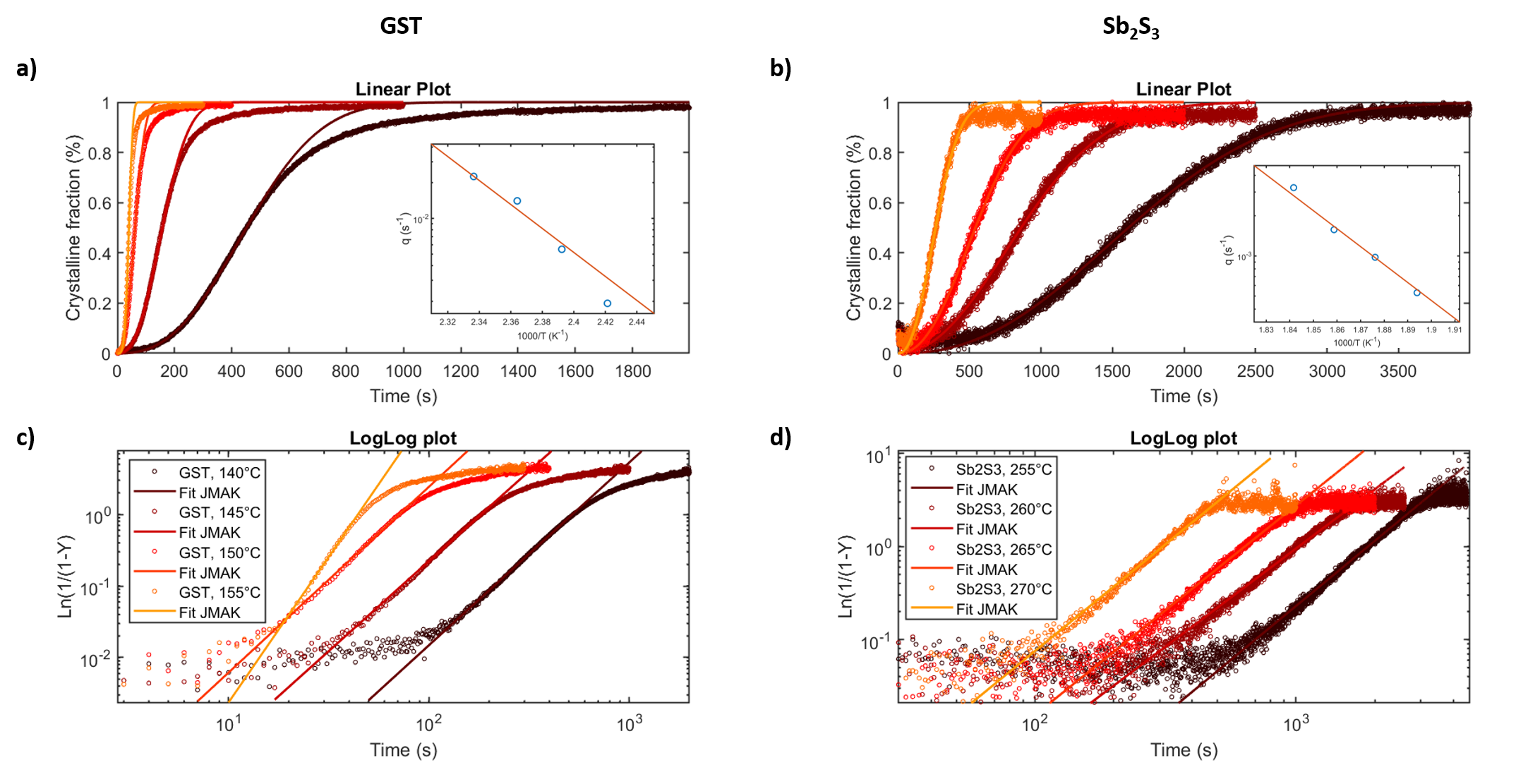}
  \caption{Crystallization fraction evolution as a function of time for three different temperatures for GST (a) linear plot, c) loglog plot and Sb$_2$S$_3$ (b) linear plot and d) loglog plot. Insets in a) and b) are Arrhenius plots of the q factor versus 1000/T.}
 \label{Figure3}
\end{figure*}

Where $Y$ is the fraction of transformed material, ranging from 0 to 1, $q$ is a time constant encompassing both the nucleation rate and the growth rate, which is a function of temperature and $m$ an integer that depends on the nature of transformation (nucleation or growth-dominated, linear or polyhedral). In our case, the fraction of transformed material represents the progressively crystallized PCMs in an amorphous host of the same material. Normalizing our previously measured values of $Is$ from 0 to 1 (i.e. fully amorphous to fully crystalline), we can use this JMAK equation to fit the experimental data and extract the values for $q$ and $m$. This procedure is valid as long as the $Is$ values linearly scale with the crystalline fraction (see the supplemental material for more information). In figure \ref{Figure3}, we show the evolution of the crystalline fraction as a function of time for different temperatures in both GST and Sb$_2$S$_3$.
As displayed in figure \ref{Figure3}, it is possible to fit a significant part of the evolution of the crystalline fraction for both PCMs at all temperatures with this formalism. Let us first focus on the GST layer. It is clear from the comparison between fits and experimental data that the JMAK model is insufficient to fully describe the whole transformation from amorphous to fully crystalline GST. This result is consistent with previous reports showing that GST-225 only partially complies with the approximations of the JMAK model \cite{senkader2004models}. Indeed, the JMAK model is only valid if the nucleation sites are homogeneously distributed in the layer and if the growth rate is constant. Two conditions that are known to not be fulfilled in GST. However, the part of the transformation that immediately follows the incubation time and until about 75$\%$ of the transformation indeed follows a JMAK evolution, with a value of $m=2.5$. By fitting this part, we extract values of $q$ that are plotted in the inset of figure\ref{Figure3}a) which are fitted with an Arrhenius law. This analysis enables the extraction of an activation energy $E_a$=2 eV that is perfectly in line with previous studies \cite{weidenhof2001laser}. 

On the other hand, the progressive crystallization of Sb$_2$S$_3$ is remarkably well fitted by the JMAK model throughout its transformation and for all studied temperatures, with a constant value of $m=2.2$. Fitting the obtained $q$ values at different temperatures with the Arrhenius law, we extract an activation energy of $E_a$=2.67 eV for the crystallization of Sb$_2$S$_3$. This result implies that Sb$_2$S$_3$ layers fully respect the approximations of the JMAK model, that is: a randomly and homogeneously distributed apparition of nucleation sites, as well as a constant nucleation and growth rate.
To the best of our knowledge, we report here for the first time that the JMAK formalism perfectly applies to the crystallization of Sb$_2$S$_3$.

We have seen that combining a careful selection of wavelength for real-time monitoring of PCMs and constant temperature for gradual crystallization not only enables precisely programming them to arbitrary crystalline fractions, but also to have deeper insights on the physics at play in the transformation. While the scope of the study reported here was to present the method and highlight its usefulness, future works should capitalize on it for several important studies. First of all, PCMs are unique not only because of their changes in physical properties upon crystallization, but also because of their reversibility. This reversibility includes cycles of crystallization-amorphization process with volume expansion and contraction \cite{raoux2009phase,weidenhof2001laser}. It is then crucial to rule out the thickness dependence from the analysis and therefore our proposed method will have a clear impact on decorrelating the thickness modulation from the real-time measurements of optical properties. Not to mention that in-situ monitoring is also mandatory when there are spatial variations of crystalline domains: different properties may arise with respect to the exact device location, because of the random apparition of nucleation sites.
There are also a large number of situations in which a real time analysis such as the one presented here is mandatory. For example, as physical properties of PCM depend on various factors such as stoichiometry, thickness, volume, substrate and capping layer, one could either or both use this method as a model-free method to program PCMs at arbitrary levels of crystallization, or to analyze the influence of these factors in the physics of the change of phase.
On a more fundamental level, the presented method could be exploited to study the fundamentals of phase change and their links to different stimuli and introduction of defects \cite{jin2022probing}.

We have presented and experimentally demonstrated a method to follow in real time the progressive crystallization of PCMs. This in-situ monitoring was first exploited to program both GST and Sb$_2$S$_3$ to arbitrary fractions of crystallization, hence enabling a robust and model-free method to encode multiple intermediate levels in the optical properties of these PCMs. Furthermore, as this transformation monitoring rules out the contributions of thickness modifications and thermo-optic effects, we exploited it for deeper analysis of the physics of phase change in these materials. By doing so, we provide for the first time direct experimental evidences that the progressive crystallization of Sb$_2$S$_3$ exactly follows the Johnson-Mehl-Avrami-Kolmogorov formalism. 
Therefore, the simple yet robust method presented here provides a very interesting pathway for both engineering devices and obtaining fundamental insights in the physics of phase transformation.

\begin{acknowledgments}
We acknowledge fundings from the French National Research Agency (ANR) under the project MetaOnDemand (ANR-20-CE24-0013). We thank F. Bentata and C. Zrounba for fruitful discussions and O. Hector for the samples preparation.
\end{acknowledgments}

\medskip

\bigskip

%%%%%%%%%%%%%%%%%%%%%%% References %%%%%%%%%%%%%%%%%%%%%%%%%

%%%%%%%%%% If using BibTeX:

\nocite{*}
\bibliography{ProgPCM}% Produces the bibliography via BibTeX.

\end{document}

\preprint{arXiv}

\title{Supplementary materials for:\\ A simple method for programming and analyzing multilevel crystallization states in phase-change materials thin films}
%\title{Programming and analyzing multilevels of partial crystallization in phase-change materials thin films}
%Alternative titles:
% A simple method for programming and analyzing partial crystallization in phase-change materials thin films
%Real-time monitoring of multilevel crystallization
% Force line breaks with \\

\author{Arnaud Taute}%
\affiliation{Université de Lyon, Institut des Nanotechnologies de Lyon (INL) UMR 5270 CNRS, École Centrale de Lyon, 36 avenue Guy de Collongue, 69134, Ecully, France}%
\affiliation{STMicroelectronics, 850 Rue Jean Monnet, Crolles, 38920, France}

\author{Sadek Al-Jibouri}%
\affiliation{Université de Lyon, Institut des Nanotechnologies de Lyon (INL) UMR 5270 CNRS, École Centrale de Lyon, 36 avenue Guy de Collongue, 69134, Ecully, France}%

\author{Capucine Laprais}%
\affiliation{Université de Lyon, Institut des Nanotechnologies de Lyon (INL) UMR 5270 CNRS, École Centrale de Lyon, 36 avenue Guy de Collongue, 69134, Ecully, France}%

\author{Stéphane Monfray}%
\affiliation{STMicroelectronics, 850 Rue Jean Monnet, Crolles, 38920, France}

\author{Xavier Letartre}%

\author{Nicolas Baboux}%

\author{Guillaume Saint-Girons}%

\author{Lotfi Berguiga}%

\author{Sébastien Cueff}%
\email{sebastien.cueff@cnrs.fr}

\affiliation{Université de Lyon, Institut des Nanotechnologies de Lyon (INL) UMR 5270 CNRS, École Centrale de Lyon, 36 avenue Guy de Collongue, 69134, Ecully, France}%

\maketitle

\section{\label{sec:level1}Ellipsometry models and fits}

Is, Ic parameters:

In the case of phase-modulated ellipsometer, like the one we use, we do not measure $\Psi$ and $\Delta$ directly. Instead, we measure functions of $\Psi$ and $\Delta$. Here, we measure Is and Ic which are defined as:

Is = sin (2$\Psi$).sin ($\Delta$),	(S1)
Ic = sin (2$\Psi$).cos ($\Delta$),	(S2)

These trigonometric functions Is and Ic are directly related to $\Psi$ and $\Delta$. They depend on the measurement conditions, in our case these above equations are only valid for the situation when analyzer is at 45° and the modulator at 0°. Throughout this work, we directly fit the theoretical models to the Is and Ic values.

Data Acquisition and Optical Modeling:
The complex dielectric functions of thin films of VO2 and their thicknesses can be derived by fitting realistic optical models to the experimental data. The acquired ellipsometric parameters Is and Ic of the thin film sample have been collected at a 70° angle over a spectrum range of 260 - 2100 nm. 
The acquisition was carried out at room temperature.

To model the complex permittivity of GST we used the Tauc-Lorentz model with XX oscillators. The capping layer is modelled using a reference dispersion (ref.). To account for the progressive crystallization, we let the Tauc-Lorentz parameters as free fit parameters.  The model is then subject to free fit using least-squares curve fitting thanks to the Levenberg-Marquardt algorithm.

\section{\label{sec:level1}Justification on the linearization of Is parameter}

1. Variation of refractive index versus crystalline fraction*
2. Variation of Is versus crystalline fraction
3. Discussion

\bigskip

%%%%%%%%%%%%%%%%%%%%%%% References %%%%%%%%%%%%%%%%%%%%%%%%%

%%%%%%%%%% If using BibTeX:

\nocite{*}
\bibliography{ProgPCM}% Produces the bibliography via BibTeX.